# Bespoke scapegoats: Scientific advisory bodies and blame avoidance in the Covid-19 pandemic and beyond

Roger Koppl[1*], Kira Pronin[1†‡], Nick Cowen[2§], Marta Podemska-Mikluch[3¶] and Pablo Paniagua Prieto[4‖]

[1]Syracuse University [2]University of Lincoln
[3]Gustavus Adolphus College
[4]King's College London

**Abstract:** Scholars have not asked why so many governments created ad hoc scientific advisory bodies (ahSABs) to address the Covid-19 pandemic instead of relying on existing public health infrastructure. We address this neglected question with an exploratory study of the US, UK, Sweden, Italy, Poland, and Uganda. Drawing on our case studies and the blame-avoidance literature, we find that ahSABs are created to excuse unpopular policies and take the blame should things go wrong. Thus, membership typically represents a narrow range of perspectives. An ahSAB is a good scapegoat because it does little to reduce government discretion and has limited ability to deflect blame back to government. Our explanation of our deviant case of Sweden, that did not create and ahSAB, reinforces our general principles. We draw the policy inference that ahSAB membership should be vetted by the legislature to ensure broad membership.

[*]E-mail: rkoppl@syr.edu
[†]Corresponding author.
[‡]E-mail: kpronin@syr.edu
[§]E-mail: nick.cowen@uclmail.net
[¶]E-mail: mpodemsk@gustavus.edu
[‖]E-mail: pablo.paniagua_prieto@kcl.ac.uk

## 1 Introduction

Why did so many governments create *ad hoc* scientific advisory bodies (ahSABs) to address the Covid-19 pandemic, instead of relying on existing public health or crisis response infrastructure? This puzzle has gone unrecognized, despite the saliency and importance of ahSABs during the pandemic. For example, in a valuable article that adds "bureaucratic demand" to the "traditional political market framework," Curley, Federman, and Shen (2023) rightly note the blame-avoidance incentive to "shift policy-making" to experts. But (given their empirical context) they overlook the

possibility of creating an ad hoc body of external experts for this purpose.[1]

Creating such bodies is a strategic decision involving tradeoffs. Large-scale, prolonged crises (Hinterleitner, Honegger, and Sager 2023) create a dilemma for sitting governments. Citizens expect the government to act, but acting is risky when outcomes are uncertain (Hood 2002, 2011). Experts provide an escape hatch. The government can blame them for unpopular policies. "We hate this policy, but the experts say it's necessary." And governments can blame them for policy failures. "We didn't fail; the experts failed."

This strategy is imperfect because delegating responsibility to experts limits the government's policy discretion (Weaver 1986), and because experts can blame the government back (Hood 2002). We argue that creating an ad hoc scientific advisory body resolves this dilemma by restoring policy discretion and reducing the risk of blame bouncing back onto the government. The current blame avoidance literature assumes that the government directs blame to political actors and experts permanently embedded in the political system (Hinterleitner, Honegger, and Sager 2023). It has not considered the possibility that the governments can *construct* a scapegoat that restores policy discretion and mitigates the risk of blame bouncing back. Our paper fills this gap.

Pamuk (2022) notes, "The scientific advisory committee is a neglected political institution whose importance became clear during the COVID-19 pandemic." Our analysis expands the understanding of this neglected institution, whose importance is growing in importance. We investigate the use of ahSABs during the Covid-19 pandemic through case studies of the USA, UK, Sweden, Italy, Poland, and Uganda. We seek to understand whether governments facing prolonged crises appoint ahSABs as 'bespoke scapegoats,' and we formulate new hypotheses about them.

Consistent with our theory, we find that governments use ahSABs as strategic instruments of blame shifting. During the Covid-19 crisis, such tactics were evident. Flinders (2020) noted at the time how leading politicians would be "flanked both figuratively and literally by 'the experts'" and intone "the golden phrase that is 'following the expert advice we are receiving'".

We also identify features of ahSABs that make them attractive blame targets. Blame sticks to them because they are compact, well defined, and easily identifiable entities with an exalted official status. Second, they can be created at will and, if necessary, replaced with a more compliant body. Third, their mandate and membership are easily controlled, enabling governments to stack ahSABs with likeminded experts. This unanimity of opinion, makes it easier to shift blame (Hood 2002). Finally, they are easily represented as an external authority that the government must obey to behave responsibly.

---
[1] Their empirical application is to state-level governments in the US, which create ahSABs infrequently.

In Section 6, we discuss institutional reforms that could mitigate these concerns.

## 2 Literature review

Weaver (1986) launched the modern blame-avoidance literature. His seminal analysis addressed the puzzle of "self-limitation of discretion by policymakers" (1986, p. 371). The standard view had been that policymakers sought to claim credit. Weaver observed, however, that politicians would often offload decision-making despite the consequent loss of discretion, doing so to avoid blame rather than take credit. Blame avoidance may, in some cases, be more important than credit claiming because of "negativity bias" among voters. Voters, he explained, tend to be "more sensitive to real or potential losses than they are to gains." Hood (2011) and James et al. (2016) also discuss negativity bias. See James and Olsen (2017) and James and Jilke (2017) for discussions of the empirical literature on negativity bias, which includes both observational and experimental studies.

Like Weaver, Hood (2002) begins his analysis with an observation. Citing Dryzek (1995), he notes the then-recent emergence of a "risk industry." He cites Douglas (1992) for his guiding principle that is there is no such thing as a blame-free risk. Thus, risk casts politicians into a "blame game" from which they cannot escape. Hood (2002) lists three strategies politicians use to avoid blame. Presentational strategies "use arguments to minimize or avoid blame." As Hood (2011) summarizes, "Spin your way out of trouble." Policy strategies pick the policy thought to minimize flak. As Hood (2011) summarizes, "Don't make contestable judgments that create losses." Agency strategies select "institutional arrangements to minimize or avoid blame, for example choosing between direct control and delegation." As Hood (2011) summarizes, "Find a scapegoat."

Hood emphasizes strategic interdependence, as suggested by his use of the word "game." He notes that delegation is one strategy of blame shifting. He also notes that experts and others to whom blame is shifted may be able to shift it back onto the politicians, which we call "biting back." Politicians wishing to blame experts therefore face a dilemma. Picking experts from only one school helps ensure unanimous conclusions, which makes it easier to shift blame to them. But that strategy makes it more likely that "the experts' conclusions or recommendations will be contested by other experts, leading to blame-sharing between politicians and groups of experts rather than a blame shift" (p. 31).

Bache et al. (2015) argue that "fuzzy governance" creates "fuzzy accountability." Fuzzy governance is the polycentrism described by Ostrom, Tiebout, and Warren (1961), Ostrom and Ostrom (1971), and Ostrom (1999). Their argument brings the theory of blame avoidance together with the theory of multi-level governance. The gist of it is that the "complex and overlapping institutional landscape" of polycentric systems makes it hard to say who should be held "accountable" when things go wrong.

Hinterleitner, Honegger, and Sager (2023) build on this idea of "fuzzy governance." They note that the prior literature on blame shifting had not addressed "large-scale prolonged crises," i.e., situations where the central governments are under pressure to act. They identify three blame targets for a national government: lower-level government units, citizens, and experts. Fuzzy governance structures make it easy to blame lower-level governments, thus sparing citizens and experts. (See also, Heinkelmann-Wild and Zangl 2020, 956; Rittberger, Schwarzenbeck, and Zangl 2017.) This strategy falls apart in large-scale prolonged crises because of expectations that the central government *act*. National governments are then more likely to shift blame to experts and citizens. It might seem obvious that experts and not citizens should be blamed. But experts may well have a better reputation than the government, in which case, "blame-shifting moves will backfire when experts push back against them."

It might seem that contracting out could be another strategy of blame shifting. The results of James et al. (2016) and Marvel and Girth (2016) suggest that contracting out to private firms is generally an ineffective blame-shifting strategy. Thus, we think it not necessary to add private enterprises to our list of blame targets.

Thus far, the literature has given little attention to how experts are organized as potential scapegoats. As far as we know, only Mavrot (2022) has considered alternative organizational structures for experts-as-scapegoats, but her classification is limited to "external" and "administrative" experts (though see Heinkelmann-Wild et al. 2023). She also implicitly assumes expert opinion is homogeneous, which is not generally true and approximated only when an expert body, such as a professional association, has an epistemic monopoly (Koppl 2018). Hinterleitner and Sager (2017) note that "the institutional design of political systems" is a strategy of "anticipatory" blame avoidance behavior. For examples, they cite Mortensen (2013b, 2013a) who considers decentralization and public sector reforms. Interestingly, ahSAB are *centralizing* and not generally considered "reforms." We seem to be the first to propose that the point of ahSABs is to be bespoke scapegoats. There are several papers investigating the use of SABs during Covid-19 (Hodges et al. 2022) as well as their membership structure (Rajan et al. 2020), but they have not recognized the pretextual nature of ahSABs.

## 3 Theory

Our theory is based on models of blame avoidance. We make the standard assumption that office holders seek to claim credit for popular policies, and avoid blame for unpopular policies. They generally prefer avoiding blame to claiming credit (Weaver 1986).

Under normal circumstances, national governments may try to avoid potential blame for risky or unpopular policies by shifting responsibility to lower-level governments. But in a large-scale, prolonged crisis, citizens expect the government to act. Attempts to diffuse responsibility to lower-level governments or other fuzzy government structures will appear as shirking responsibility. In such circumstances, governments may try to shift blame to citizens or experts.

Two considerations discourage governments from using experts as scapegoats despite the obvious perils of blaming the people. First, experts may limit the government's discretion. Second, if blamed for a bad outcome, experts may bite back with criticisms of the government. This puts the government is in a bind and might induce it to scapegoat the people for want of a better option (Hinterleitner, Honegger, and Sager 2023). Creating an ad hoc scientific advisory body (ahSAB) gets the government out of this bind. An ahSAB gives the government policy discretion *and* a scapegoat ill-equipped to bite back.

ahSABs have four properties that make them good blame targets.

First, they are compact, well-defined, and easily identifiable entities with an exalted official status. Therefore, blame will stick to them relatively easily.

Second, because they are ad hoc, they may be created and disbanded at will. Thus, it is relatively easy to ensure that the ahSAB will not bite back or propose policies undesired by the government. The members of an ahSAB have an incentive to be docile, which largely frees the government to act on its discretion.

Third, the ahSAB's membership is easily controlled, ensuring that the ahSAB will not bite back but, instead, support the government. Because membership is easily controlled the SAB can be composed entirely of experts who agree with one another. Such unanimity of opinion, Hood (2002) explains, makes it easier to shift blame to the SAB.

Finally, the government's ahSAB is composed of experts and is thus easily represented as an external authority that the government must obey if it to behave responsibly. Angelou et al. (2023) discuss the literature on "evidence-informed policy," which supports the generalization that "the involvement of experts in pandemic management increased citizens' trust" in general and in the context of the Covid pandemic (2023). Their survey evidence gives further support to the proposition. See also Ladi et al. (2021) and Robinson et al. (2021). Thus, scientific expertise is not only useful for implementation of effective policy but also an attractive thing for policymakers to counterfeit.

Thus, ahSABs resolve the puzzle with which we began: Why did so many governments create ad hoc scientific advisory bodies (SABs) to address the Covid-19 pandemic? Our theory establishes the clear presumption that national governments will create ahSABs to help them during prolonged crises, that the main help they provide is being good scapegoats, and that, ahSABs are generally pretextual. In other words, they are political, and not epistemic.

While the advantages ahSABs in a large-scale prolonged crisis are great, they do not eliminate the "politicians' dilemma" Hood (2002, p. 31, n. 50) identified. The narrower the membership, the *less* likely is bite back *from within* the SAB, but the *more* likely is bite back *from without* the SAB as excluded experts criticize the government's policy or its choice of SAB membership. The risk of bite-back from within an SAB discourages governments from including stakeholders in an ahSAB. And, indeed, many governments excluded them from Covid-19 ahSABs. The risk of bite-back from without an ahSAB may not always be great. When it is, governments have counterstrategies to delegitimize dissenting opinions. And, indeed, some governments were active participants in controversies over the legitimacy of divergent opinions on crucial scientific issues of the Covid-19 crisis. Further exploration of such counterstrategies is beyond the scope of our study. We hope to return to the theme in later works.

Figure 3.1 and Table 3.1 summarize our argument:

*Table 3.1. Properties of SABs that make them a good blame target during prolonged crises*

| Feature | Mechanism of action |
|---|---|
| Ad hoc | Mandate (agenda) controlled by government. Can create/disband and replace with another (with some political cost) if necessary. Less likely to bite back than permanent political actors or experts in general. |
| Compact, identifiable, highly visible, official | Blame will stick more easily because voters will perceive the SAB is responsible for the advice. |
| Membership controlled by government | Will not propose anything that would make government look bad. Government maintains policy discretion. |
| Filled with independent experts | Creates an aura of legitimacy |

*Figure 3.1. Main framework: Government's decision-making in a large-scale, prolonged crisis*

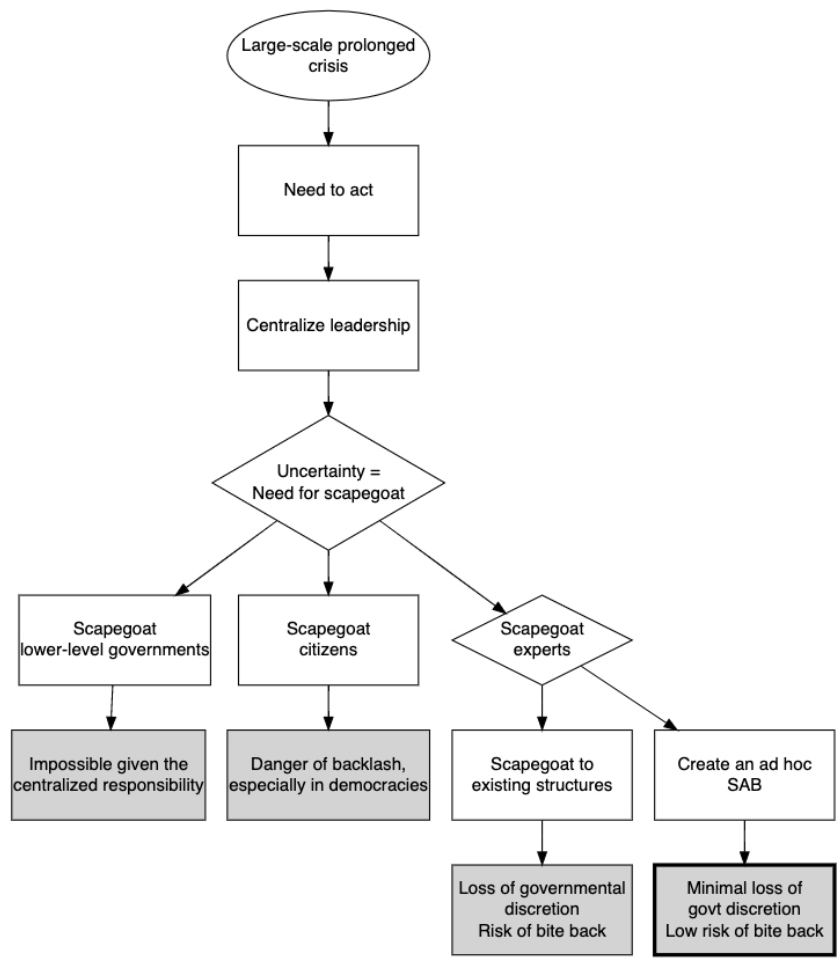

# 4 Research design and data

We conduct case studies of six countries (USA, UK, Sweden, Italy, Poland, and Uganda), focusing on the main SABs appointed or deployed during the Covid-19 pandemic. Our cases provide variation in potential independent and intervening variables in terms of geographical location and political systems. We include a deviant case where the outcome variable (appointment of an ahSAB) differs from the other cases: in Sweden, the government did not appoint a Covid-19 ahSAB, and agenda-setting on Covid-19 policy was largely delegated to a government agency. Our design is exploratory (Gerring 2006, p. 65). That is, we use our cases to identify new hypotheses about SABs.

We used publicly available primary (government reports, executive orders) and secondary sources (published case studies, news, etc.) to construct the case narratives. We present our findings in a cross-case fashion, focusing on the main predictions of our theory and additional insights we have discovered.

Table 4.1 lists the SABs included in our study. Additional information can be found in the in the Appendix.

*Table 4.1 List of SABS included in the case study*

| Country | SAB |
|---|---|
| *Italy* | CTS (Comitato Tecnico-Scientifico) |
| | CES (Comitato di esperti in materia economica e sociale) |
| | ISS (Istituto Superiore di Sanità) |
| | CNB (Comitato Nazionale per la Bioetica). |
| *UK* | SAGE (Scientific Advisory Group for Emergencies) |
| | NERVTAG (The New and Emerging Respiratory Virus Threats Advisory Group) |
| | JCVI (The Joint Committee on Vaccination and Immunisation) |
| *USA* | White House Coronavirus Task Force (Trump administration) |
| | Covid-19 Advisory Board (Biden administration) |
| | White House Covid-19 Response Team (Biden administration) |
| | Presidential Covid-19 Health Equity Task Force (Biden administration) |
| *Poland* | Medical Council (Rada Medyczna) |
| | Council for COVID-19 (Rada ds. COVID-19) |
| | The Crisis Headquarters (Sztab Kryzysowy) |
| *Sweden* | PHAS (Public Health Agency of Sweden) |
| *Uganda* | National Task Force (NTF) |
| | Parliament's Covid-19 task force |
| | Regional parliamentary task force teams (RTF) |

# 5 Cross-case comparison and discussion of results

## 5.1 Reliance on ahSABs vs permanent government structures; appointment of ahSABs and the timeline of the pandemic

With the exception of Sweden, the governments in our sample countries made use of ahSABs. "In Italy," Camporesi et al. (2022) explain, "the management of the pandemic adopted a classic top-down approach." Camporeis et al. (2022) identify the "key" expert committees: 1) *the CTS: Technical and Scientific Committee* (Comitato Tecnico-Scientifico), 2) *the CES: Economic and Social Committee* (Comitato di esperti in materia economica e sociale), 3) *the ISS: the Italian National Institute of Health* (Istituto Superiore di Sanità), 4) *the CNB*, which they describe as "Italy's main public health research center," and the national bioethics committee (Comitato Nazionale per la Bioetica). Camporesi et al. (2022) say, "During the first wave in the spring of 2020 the main advisory body for the production of expert advice in Italy was the Technical and Scientific Committee (CTS), set up by the Head of the Civil Protection Department (Ordinance n. 630, Civil Protection Department, 2020a)."

In the United Kingdom, the key SABs involved in the Covid-19 policy were *the Scientific Advisory Group for Emergencies* (SAGE), *the New and Emerging Respiratory Virus Threats Advisory Group* (NERVTAG) and, *the Joint Committee on Vaccination and Immunisation* (JCVI). Of these, SAGE was the most prominent and influential. While SAGE committees are an established feature of British government, each instance is activated by the Prime Minister on an *ad hoc basis with no formal criteria for membership, which can change from meeting to meeting*.

In the United States, the Covid-19 pandemic spanned two presidential terms. On January 29, 2020, President Trump appointed *the White House Coronavirus Task Force,* consisting of a chair, a Covid-19 coordinator reporting to the President (the coronavirus 'czar'), and 23 subject matter experts from the White House and several government agencies. In early May 2020, President Trump proposed that the Task Force be phased out to accommodate another group centered on reopening the economy. After a backlash, he stated that the coronavirus task force would continue "indefinitely". However, by the end of May, the task force met only once per week, whereas it had earlier met every day of the week. It existed until the end of the presidential term.

On November 9, 2020, before naming any White House staff or cabinet appointments, President Biden announced the appointment of a transitory *White House Covid-19 Advisory Board* composed of 16 health experts. The board was dissolved on January 20, 2021, after President Biden was sworn in, and was replaced by *the White House Covid-19 Response Team* (Executive Order 1398). The new team included a Covid–19 Response Coordinator, a Deputy Coordinator and 13 members, among whom were medical professionals, government officials and experts on communication. The Covid–19 Response Coordinator reported directly to the President and advised and assisted the President and executive departments and agencies in the Covid–19 response.

On November 10, 2020, the Biden administration also established the *Presidential Covid-19 Health Equity Task Force* (Executive order 13995), whose mission was to provide recommendations for mitigating inequities caused or exacerbated by the Covid-19 pandemic. The Task Force included a chair, and six ex-officio members from Federal agencies and other members with diverse backgrounds and expertise.

Like several other Eastern European states, Poland *did not initially establish a dedicated ahSAB*. During the first wave, decisions were carried out within established institutions, the key among them being *the Government Crisis Management Team* (established in 2007). On March 9th, the government introduced health checks at the borders. The following day, a ban was imposed on gatherings of more than 1,000 people in open spaces and more than 500 people in confined spaces. On March 12th, the first death from coronavirus was recorded in Poland, prompting the closure of schools, cultural institutions, and universities, later replaced by a shift to remote learning. Each day, additional bans were added: prohibition on all public gatherings, closure of banks, boulevards, and beaches, closure of all non-essential shops, suspension of all sporting events and activities, mandatory 14-day quarantine for travelers, limitations on the use of public transportation, and mandatory wearing of face masks in public places, including shops and public transportation. By April 10, the stringency of the lockdown in Poland approached that of Italy (Hale et al. 2021). Like the UK, among others, Poland subsequently fell into on-again-off-again restrictions. It was on-again when case counts climbed, but off-again for the election in June 2020 and for Christmas 2020.

In July 20th 2020, the Minister of Health established a Covid-19 group that included health experts and representatives of several state agencies (Badora-Musiał and Dusza 2021). In November 2020, the Prime Minister established the *Medical Council for Covid-19* tasked with developing recommendations on anti-Covid-19 measures. *On January 14, 2022, 13 out of 17 members of the Council s resigned, protesting that their recommendations were being ignored and* that the government officials were undermining the importance of vaccinations. A week later, the Prime Minister abolished the Medical Council and established a new Council for Covid-19 on the same day. The new council's mandate was expanded to include the analysis and assessments of the economic and social conditions, in addition to the health situation. With this expansion, came a larger, more diverse membership The 17 members of the original Medical Council were primarily medical experts, the new Council for COVID-19 had 29 members that, besides medical practitioners, included socio-economic experts, institutional representatives, and researchers. This episode perfectly illustrates how ahSABs were used politically. Trouble with the old ahSAB led its speedy dissolution and replacement with another ahSAB having a more compliant membership.

In contrast to other countries, Sweden *relied on its existing institutions to provide pandemic-related expertise, instead of appointing special ad hoc bodies*. *The Public Health Agency of Sweden*, or *PHAS* (Swedish: *Folkhälsomyndigheten*) was the main provider of expert advice about the Covid-19 pandemic, with the state epidemiologist Anders Tegnell playing a key role in shaping the response strategy. Another player in the Covid-19 response was *the National Board of Health and Welfare*, or *NBHW* (Swedish: *Socialstyrelsen*), which was responsible for ensuring health care capacity during the pandemic. It issued recommendations about various topics, such as end-of-life-care of Covid-19 patients (Olofsson and Vilhelmsson 2022). Finally, Sweden's 21 regional councils (including 21 county administrative boards) and 290 municipalities bore the responsibility for planning and implementing the local response to the pandemic in light of the national recommendations (Tegnell 2021).

Sweden also differed from other countries in that it did not declare a state of national emergency or implement lockdowns, mask mandates, closures of daycares and primary schools, or stay-at-home orders (Kavaliunas et al. 2020; Mens et al. 2021; Nilson 2021; Pierre 2020). Instead, citizens learned about voluntary self-protection measures from the website of the PHAS and frequent press conferences held by Anders Tegnell and Prime Minister Stefan Löfven (Giritli Nygren and Olofsson 2020). As the pandemic progressed, the Government took on a more central role. In 2020 the *Riksdag* introduced several legally binding regulations with limits on public gatherings, restaurant operations, and visits to elder care. In January 2021, the provisional Covid-19 Act [2021: 4] added a series of further restrictions with concomitant enforcement instruments (Government Offices of Sweden, 2021a, 2021b, 2021c, 2021d), which expired in September 2021 (Lynggaard, Jensen, and Kluth 2023).

When Covid-19 first arrived in Uganda in early March, 2020, the government took a series of measures to minimize the risk of transmission, including closing entry points into the country, banning public gatherings and the use of public transport, closing schools and places of worship, and declaring a national lockdown and curfew. In June 2020, the National Security Council set up a multisectoral *National Task Force* (NTF) to coordinate with the government to fight against the spread of coronavirus disease in communities. The NTF was organized under the President and chaired by the Prime Minister.

On June 29, 2020, the Deputy Speaker Anita Among announced the appointment of the 40-member *Parliament's Covid-19 task force* in a press conference at the Parliament. The task force was charged with representing the parliament to the NTF and coordinating the overall interventions of the four *regional parliamentary task force teams* (RTF), which were announced on the same day. The regional teams were given two weeks to conduct sensitization and awareness programs on mass media to educate the population about Covid-19 and to assess the role of private sector healthcare providers in Covid-19 management challenges. Their mission included assessing the operation, administration and management of funds and other resources appropriated for them in the management of the pandemic. The task forces were also expected to carry out field visits and assess the state of health care systems in regional referral hospitals and districts. Their terms of reference required them to regularly brief the speaker of parliament and the parliament task force on the national response as well as submitting a report to parliament for debate.

**5.2 ahSAB membership composition**

The composition of the main Covid-19 ahSABs tended to be overwhelmingly drawn from public health and infectious disease. For Italy, there was "*a predominance of epidemiologists and infectious disease specialists over social scientists in the mobilization of experts for the management of the crisis in Italy*" (Camporesi, Angeli, and Fabbro 2022, p. 1). Also, "On April 10th, 2020, the Prime Minister Giuseppe Conte set up a Council of experts on economic and social matters (also known as . . . the 'Colao Committee' from the name of its chair, Vittorio Colao) with the mandate to investigate the impact of the pandemic on socioeconomic activities, and to provide key recommendations on how to support the Italian social and economic recovery" (Camporesi et al. 2022). Their recommendations, Camporesi et al. (2022) report "did not have a direct impact on national decrees, in stark contrast with the advice provided by the CTS."

In the UK, the ahSAB composition was skewed towards experts in epidemiology, public health, and other medical specialties. Academic statisticians and biologists were also represented. This focus is also reflected in the issues they explored and the questions asked of them by the Government, namely what the impact of different scenarios and policy responses would be on the spread of the virus and the impact on healthcare services (Jarman et al. 2022). There were exceptions. For example, Prof. Rebecca Allen, an expert in education with a research focus on the impact of government policy on experiences of teaching in schools, was a SAGE advisor. Prof. Robert Dingwall, a consulting sociologist and professor at Nottingham Trent University, served on both NERVTAG and the JCVI. He became prominent in the UK in May 2020 for disclosing the emerging expert consensus that outdoor transmission of Covid-19 was negligible.

In the U.S., the Trump administration's White House Coronavirus Task Force had a wider sectoral representation than the Biden administration's White House Covid-19 Response Team. It included six ex-officio members from Federal agencies (Departments of Agriculture, Education, Health and Human Services, Housing and Urban Development, Justice, and Labor), On February 26, 2020, U.S. vice president Mike Pence was named to chair the task force, and Deborah Birx was named the Covid-19 coordinator, and on May 15, 2020, additional members with expertise with vaccines, therapeutics and worker safety were added. By comparison President Biden's White House Covid-19 Response Team included a Covid–19 Response Coordinator (Jeffrey Zients), a Deputy Coordinator (Natalie Quillian) and 13 members, among which were medical professionals, government officials and experts on communication. The Response Team was not diverse and, in particular, lacked anyone with an economics background.

The most diverse of all the ahSABs in our sample was President Biden's *Presidential Covid-19 Health Equity Task Force*. Its members represented a diversity of backgrounds and expertise, a range of racial and ethnic groups, and representatives of a number of other important populations, such as children and youth; educators and students; health care providers, immigrants; individuals with disabilities; LGBTQ+ individuals; public health experts; rural communities; state, local, territorial, and Tribal governments; and unions. However, it had a much narrower mandate than the other ahSABs and did not have a large impact on the pandemic.

In Poland, there was no independent SAB advising the Prime Minister or Minister of Health during the first wave. The PM and the Minister of Health frequently referenced consultations with experts in their public appearances and social media posts. But *the names of the experts remained anonymous throughout the first wave of the pandemic*. Eventually, on May 20, the Ministry of Health published a list of its consultants, comprised of the employees of the National Institute of Public Health - National Institute of Hygiene, and academics with backgrounds in mathematics and statistics.

From September 2020, two formal SABs were in place. The first team focused on monitoring and forecasting the epidemic, providing reports, and proposing improvements in data reporting. It was led by the director of the National Institute of Public Health - National Institute of Hygiene (NIZP-PZH), with Paweł Maryniak, the deputy director of the Department of Analysis and Strategy in the Ministry of Health, serving as his deputy. The team consisted of representatives from the Ministry of Health, NIZP-PZH, the Chief Sanitary Inspectorate (GIS), the Central Statistical Office (GUS), the Agency for Health Technology Assessment and Tariffication (AOTMiT), and the e-Health Center.

The second team, Headquarters (Sztab Kryzysowy), was responsible for analyzing the current epidemiological situation in Poland and other countries, establishing solutions for prevention, counteraction, and combating Covid-19, and continuing the actions initiated by the team responsible for developing a strategy for combating the Covid-19 epidemic. Its members included the Chief Sanitary Inspector, the President of the National Health Fund (NFZ), the Director of the National Institute of Public Health - National Institute of Hygiene (NIZP-PZH), a national consultant in the field of infectious diseases, and representatives from various organizational units of the Ministry of Health.

From November 2020 until January 2022, the Medical Council for Covid-19 was comprised of 15 (later 17) experts in epidemiology and infectious diseases. After the resignation of 13 of the members, the reconstituted Council for Covid-19 was more diverse, but probably less independent. Its 29 members included socio-economic experts, institutional representatives, researchers, and medical practitioners. Socio-economic experts and institutional representatives were added so that the council could reconcile medical arguments with other "freedoms." In the words of the prime minister, explaining the addition of non-medical experts: "there is a possible conflict between, on the one hand, medical reasons, health care reasons, which have a common denominator called truth (…), and on the other side, expectations, which have a common denominator of a nature, or under the name of freedom".

In Uganda, the National Task Force included representatives from the Office of the Prime Minister's Office and the following ministries: Health, Internal Affairs, Defence, Works and Transport, and Trade and Industry, as well as representatives of the information and communications technology sectors, Kampala Capital City Authority (KCCA), and the private sector. The 40-member Parliament's Covid-19 task force, which represented the parliament to the NTF, consisted of MPs from various parties and regions. (RTF), The regional parliamentary task forces consisted of MPs and were led by Dr Michael Bukenya (Bukuya) who also headed the regional team for central Uganda, Dr Charles Ayume (Koboko Municipality), heading the team from northern Uganda, Dr Joseph Ruyonga (Hoima West Division), heading the western Uganda regional task force team, and Dr Emmanuel Otaala (West Budama South), leading the eastern task force.

## 5.5 Lessons

Our exploratory study gives empirical support to several hypotheses that emerge naturally from our theory.

1) ***Governments tend to use ahSABs as cover for policies they desire independently of scientific advice.***

This hypothesis is illustrated by the government of Uganda, which used the pretext of the pandemic to oppress its political opponents. It is, in part, the existence of pre-existing preferences that explains Weaver's (1986) observation that governments are reluctant to give up policy discretion. Despite the apparent diversity and checks and balances in Uganda's expert bodies appointed in response to the pandemic, and that task forces were broadly representative, both across sectors and geographically, and the parliamentary task force provided oversight of both the National Task Force and the regional task forces, there is again evidence of ahSABs – or at least information provided by them – being used politically: the government used the Covid-19 public health interventions as cover to engage in physical repression in areas that had a high concentration of the President's opponents (Grasse et al. 2021). This case shows how ahSABs are used to serve pre-existing political purposes.

Our deviant case of Sweden supports the hypothesis that ahSABs are used as cover. Recall that, in our theory, an essential function of an ahSAB is to give the government policy discretion. The unique circumstances of the "January agreement" (explained presently) so tightly bound the Löfven to a rigid set of policy preferences that it could not benefit from policy discretion.

The first Löfven government fell in September of 2018 and it proved difficult to assemble a winning coalition to replace it. Finally, in January 2019 a new Löfven coalition was voted into power. The coalition was composed of only Löfven's Social Democrat Party and the Green Party. To get themselves voted in, the coalition was driven to strike a deal with the Center Party and the Liberals. The resulting January Agreement (*Januariavtalet*) was signed on 19 January 2019 by representatives of those four parties. The agreement bound the government to a 73-point program that included strongly liberal elements including a call for "better conditions for businesses and entrepreneurs" that specifically required changes in the tax laws to benefit "growing SMEs." Lockdown would have been inconsistent with defending and supporting businesses, entrepreneurs, and growing SMEs. Löfven's red-green coalition was so fragile that any deviation from the requirements of the January Agreement would have sunk it. The end of the coalition came in July of 2021 when it tried to honor the Agreement's call for rent reform by proposing a liberalization measure that would have given landlords greater freedom to increase rents. The socialist Left Party could not abide this measure and called for a vote of no confidence, and the government fell.[2] This end to the "Löfven II" government reflects both the rigidity of the January Agreement – it put the government in a straightjacket – and the fragility of the governing coalition.

Sweden's Covid-19-era Red-Green government was, then, a weak coalition government comprising many heterogeneous interests, and which had made a pact (*Januariavtalet*) with Liberals on preserving free-market features of the economy. If the sitting government had produced an ad hoc body representing only a few narrow interests, the coalition would probably have collapsed because the interests of some party in the Swedish legislature would not have been represented. If, instead, it had created an ad hoc advisory body with sufficient breadth of membership to prevent that outcome, it could not have predicted what advice would emerge for such a body. The Swedish Public Health Authority, which took charge of the pandemic from day one, with the charismatic Chief Epidemiologist Anders Tegnell at the helm (as well as Sweden's regions and municipalities), also offered a convenient scapegoat in case of Sweden's highly unusual strategy of no mandatory lockdowns turned out to be a failure. Tegnell was also a political actor who had the ability to bite back, and overriding him would have presented difficulties. In principle, it would have required major legislative action because of provisions in Sweden's constitution that forbid ministerial rule of governmental agencies (Jonung and Hanke 2020, Bylund and Pakard 2021). The highly decentralized nature of Swedish emergency response infrastructure would have presented further obstacles. Thus, it was not in the political interest of the Swedish government to create an ad hoc emergency advisory body. Instead, Sweden simply relied on pre-existing administrative structures. Andersson and Aylott (2020) also discuss the weakness of the government as a contributing factor.

---

[2] https://www.idu.org/what-is-going-on-in-sweden-a-political-update-september-2021/

2) *The risk of "bite back" from within an ahSAB is relatively low*

We have seen that the ahSAB changed with administrations in the US. If members are chosen to be compliant, they will be relatively unlikely to voice dissent. And if they do, as we have seen with Poland's Medical Council, the dissenting ahSAB can be dissolved and replaced. *The case of UK also illustrates how sticky blame can be when it is attached to an ahSAB.* For example, the UK's main Covid SAB, SAGE, has continued to received credit or blame for the government's policy choices. An article in *Nature* (Naddaf 2023) on Angela McLean's appointment the UK's chief scientific advisor notes her role in SAGE during the pandemic, which helped "to guide the government's response to the pandemic." Quoting a prominent "science-policy researcher," the article says, "McLean's experience serving on SAGE 'could put her in a really good position' to think about reforming the body." *Thus, blame and credit are both still sticking fast to SAGE.*

3) *ahSABs are political not epistemic.*

In the United States, the Trump administration's White House Coronavirus Task Force was immediately replaced *with an interim task force* during President Biden's transition, *before even naming any White House staff or cabinet appointments.* This perfectly illustrates the intensely political nature of ahSABs. There was a great difference in composition between the Trump administration's White House Coronavirus Task Force and the Biden administration's White House Covid-19 Response Team. Even the change in name suggests that these ahSABs were never meant to provide neutral scientific advice. If an ahSAB provided neutral, objective scientific advice, a change in regime would not imply a change in ahSAB or its composition.

*4) ahSABs tend to have narrow memberships that exclude broad stakeholder interests.*

Italy's CTS has been criticized for having a "narrow range of expertise" (Pistoi 2021). Italy is a remarkable case of "near-complete overlap of technical advice and political response in the first phase of the pandemic in Spring 2020, with a key role played by the advice provided by the Technical and Scientific Committee (CTS)" (ibid., p. 1). In particular, there was "a predominance of epidemiologists and infectious disease specialists over social scientists in the mobilization of experts for the management of the crisis in Italy" (ibid., p. 1). Grizzle, Goodin, and Robinson (2020) have emphasized the importance of diversity, albeit in the slightly different context of administrative networks. Koppl (2018) emphasized "epistemic diversity." And Koppl (2023) emphasized the "diversity of inconsistent interests." If ahSABs are appointed for strategic reasons, then they will mostly consist of experts, especially of experts with similar opinions, with little or no diversity in school of thought, geography, or discipline. This hypothesis is illustrated by the difference in membership between the UK's SAGE, which is an ahSAB and its Joint Committee on Vaccination and Immunization (JCVI), which is a fully regularized standing body. As we have seen, SAGE membership was predominantly epidemiologists and experts in closely related medical fields. JCVI, by contrast, maintains a wider variety of scientific perspectives in its membership as well as lay persons. JCVI is less predictable, less likely to give unanimous opinions, and less likely to support government policy. It was, for example, "largely opposed" to vaccinating children under 16 against Covid.[3]

*5) Governments act to control the ahSABs they create.*

The British and Polish cases illustrate this hypothesis. In Poland, a recalcitrant ahSAB was simply disbanded and replaced. In the UK, "observers and government officials" attended almost all meetings of the Covid SAGE, including its first meeting on 22 January 2020. And beginning with "SAGE9," which met 20 February 2020, these regularly included personnel from No 10. An addendum to the minutes of the first Covid SAGE meeting explains that these "attendees" are there to "listen" and "to provide the scientific experts with context on the work of government

---

[3] https://amp.theguardian.com/society/2021/aug/07/jcvi-largely-opposed-to-covid-vaccination-for-children-under-16

where appropriate."[4] "Listening" and providing "context" were government levers used to control SAGE. Our deviant case, Sweden, illustrates the complementary point that scientific advice is harder to control when it comes from a pre-existing governmental body such as Sweden's PHA than when it comes from an ahSAB.

# 6 Conclusions and policy implications

Our results have policy implications for the institutional design of SABs. Especially during prolonged, large-scale crises, ahSABs are vulnerable to blame and can be a source of expert failure. This is because their appointment, mandate, and membership are controlled by the government, which has an incentive to be seen to act but also to avoid being blamed. We agree with Pamuk (2022) that the inherent tension between public interest and government incentives in the use of SABs should be mitigated by broader democratic scrutiny. Without explicit regulations or norms about their appointment and composition, governments may use ahSABs to add a veneer of objectivity and necessity to policies that have already been decided. Indeed, our cases show that the composition of Covid-19 ahSABs was not generally diverse, and the executive maintained tight control over their appointment, mandate, and member selection. The only exceptions were ahSABs whose recommendations would be predictable and would align with the government's prior policy preferences.

Based on these considerations, we suggest that there should be formal barriers to the formation of ahSABs, and governments should instead rely on the existing crisis response infrastructure. If such an infrastructure is lacking, there should be mechanism to ensure broad stakeholder representation on ahSABs. Grizzle, Goodin, and Robinson (2020) propose "deliberate recruitment" of diverse parties to administrative networks. They note, however, the difficulty posed by the "tendency of people to seek similar partners" (homophily), which works in the opposite direction. For democratic governments forming ahSABs, direct recruitment can be supplemented by a legal requirement that membership be vetted with the legislature as Koppl (2023, pp. 119-120) has proposed. This measure would help to ensure that broad stakeholder interests would be represented on the ahSAB.

---

[4] https://assets.publishing.service.gov.uk/government/uploads/system/uploads/attachment_data/file/1058442/S0369_Precautionary_SAGE_meeting_on_Wuhan_Coronavirus__WN-CoV____1_.pdf